\documentclass[%
reprint,
 amsmath,amssymb,
 aps,
]{revtex4-2}

\usepackage{graphicx}
\usepackage{dcolumn}
\usepackage{bm}

\begin{document}

\preprint{APS/123-QED}

\title{Beyond Borel Windows:\\
A Systematic Optimization Framework for QCD Sum Rules}

\author{Raphael M. Albuquerque}
 \email{raphael.albuquerque@uerj.br}
 \affiliation{Faculty of Technology\\
 Rio de Janeiro State University - UERJ, Brazil}

\date{\today}

\begin{abstract}
We propose a systematic optimization framework for determining the sum rule 
window in QCD Laplace sum rules. Instead of relying primarily on fixed 
convergence percentages and visual plateau selection, the procedure 
combines an OPE entropy criterion, local mass stationarity, and a 
correlated analysis of the Laplace parameter $\tau$, the continuum 
threshold $t_c$, and the interpolating-current mixing angle $\theta$. 
The normalized OPE entropy is introduced to quantify the redistribution 
of the QCD contributions between the perturbative and nonperturbative sectors 
and to delimit an initial Working Region. This domain is subsequently 
refined by minimizing the residual variation of the mass estimator and 
requiring weak sensitivity to the continuum threshold and to the 
current composition. As an application, we study the lowest fully 
charmed tetraquark state with quantum numbers $J^{PC}=2^{++}$ using a 
mixed diquark--antidiquark and meson--meson interpolating current. 
The optimization selects $\theta=17.2^\circ\pm0.4^\circ$, 
$\sqrt{t_c}=7.08\pm0.08~\mathrm{GeV}$, and 
$\tau=0.410\pm0.010~\mathrm{GeV}^{-2}$, leading to the mass prediction
\(
M_{c\bar c c\bar c}^{2^{++}}=(6.32\pm0.11)~\mathrm{GeV}.
\)
The predicted state lies in the near-threshold region of the 
di-$J/\psi$ spectrum and is compatible, within uncertainties,  
with the lowest resonant component reported by ATLAS and with the 
enhancement parametrized as \(BW_0\) in the recent CMS publication. 
The proposed framework provides a reproducible way of identifying finite 
domains of reduced auxiliary-parameter sensitivity and of incorporating 
the remaining dependence into the final uncertainty.
\end{abstract}

\maketitle


\section{\label{sec:chap1}Introduction}

Quantum chromodynamics (QCD) provides the fundamental 
description of strong interactions in terms of quark 
and gluon degrees of freedom. At hadronic energy scales, however, 
the growth of the strong running coupling ($\alpha_s$) and the 
phenomenon of confinement prevent a direct treatment of bound 
states through ordinary perturbation theory. The QCD sum rules 
method, originally developed by Shifman, Vainshtein, and Zakharov 
\cite{SVZ1979a,SVZ1979b,SVZ1979c}, 
offers a bridge between QCD theory and hadronic phenomenology by 
combining analyticity, dispersion relations, the operator product 
expansion (OPE), and a parametrization of the hadronic spectral
function. Since their formulation, they have been extensively 
applied to conventional mesons and baryons, heavy-quark systems
glueballs, hybrids and multiquark candidates, including tetraquark 
and pentaquark configurations
\cite{Reinders1985, Colangelo:2000dp, Narison2004, 
Narison2015,Khodjamirian:2001xq, Narison:1994ag, Jamin:2001fw, 
Ioffe:1981kw, Chung:1981cc, Leinweber:1995fn, Novikov:1979va, 
Narison:1988ts, Bagan:1989vm, Harnett:2008cw, Govaerts:1984hc, 
Govaerts:1985fx, Govaerts:1986pp, Nielsen:2009uh, 
Albuquerque:2018jkn, Albuquerque:2009ak, 
Albuquerque:2011ix, Albuquerque:2012rq, Matheus:2006xi, 
Lee:2007gs, Matheus:2004gx, Navarra:2005rq, Albuquerque:2013hua, 
Chen:2016qju}.

The method starts from a correlation function constructed with 
interpolating currents carrying the quantum numbers of the hadronic 
channel under investigation. 
For the tensor channel considered in this work as an example, 
the starting point is the rank-four two-point correlation function
\begin{equation}
\Pi_{\mu\nu,\rho\sigma}(q)
=
i\int d^4x\,e^{iq\cdot x}
\langle 0|
T\left\{
J_{\mu\nu}(x)J_{\rho\sigma}^{\dagger}(0)
\right\}
|0\rangle ,
\label{eq:two_point_correlator}
\end{equation}
where $J_{\mu\nu}$ is an interpolating current with the appropriate 
flavor content and a nonvanishing coupling to tensorial states. 
Since the correlator generally contains contributions from 
different spin components, the invariant amplitude is isolated 
through the corresponding tensor projector,
\begin{equation}
\Pi_{\mu\nu,\rho\sigma}(q)
~=~
\mathcal{P}^{(T)}_{\mu\nu,\rho\sigma}(q)\,
\Pi(q^2)
~+~ \dots
\label{eq:tensor_decomposition}
\end{equation}
where the projector is transverse, symmetric under the exchanges
\(\mu\leftrightarrow\nu\) and \(\rho\leftrightarrow\sigma\), and traceless in both index pairs, thereby isolating the \(J^{PC}=2^{++}\):
\begin{eqnarray*}
\mathcal P^{(T)}_{\mu\nu,\rho\sigma}(q)
&=&
\frac{1}{2}
\left[
\Delta_{\mu\rho}\Delta_{\nu\sigma}
+
\Delta_{\mu\sigma}\Delta_{\nu\rho}
\right]
-
\frac{1}{3}
\Delta_{\mu\nu}\Delta_{\rho\sigma},\\
\Delta_{\mu\nu}
&=&
-g_{\mu\nu}
+
\frac{q_\mu q_\nu}{q^2}.
\end{eqnarray*}
The invariant amplitude $\Pi(q^2)$ admits complementary QCD and
hadronic representations. At large spacelike momenta,
$Q^2=-q^2\gg\Lambda_{\mathrm{QCD}}^2$, its short-distance behavior is
described by the OPE,
\begin{equation}
\Pi^{\mathrm{OPE}}(q^2)
=
\Pi_{\mathrm{pert}}(q^2)
+
\sum_d
C_d(q^2)\,
\langle O_d\rangle,
\label{eq:ope_correlator}
\end{equation}
where $C_d(q^2)$ are the Wilson coefficients and $\langle O_d\rangle$
denote vacuum condensates of dimension $d$. The same QCD representation
can be expressed by a dispersion relation,
\begin{equation}
\Pi^{\mathrm{OPE}}(q^2)
=
\int_{t_q}^{\infty}
\!\!ds\,
\frac{\rho^{\mathrm{OPE}}(s)}
{s-q^2-i\epsilon}
+
\text{subtraction terms},
\label{eq:ope_dispersion_relation}
\end{equation}
with
\begin{equation}
\rho^{\mathrm{OPE}}(s)
=
\frac{1}{\pi}
\operatorname{Im}
\Pi^{\mathrm{OPE}}(s+i\epsilon),
\label{eq:ope_spectral_density}
\end{equation}
where $t_q$ is the QCD production threshold. To suppress the 
subtraction terms and reduce the relative importance of
the high-energy region, we apply the inverse Laplace, or equivalently
Borel--Laplace, transformation to both representations of the correlator.
Introducing the Laplace parameter $\tau\equiv 1/M_B^2$, the QCD side
becomes
\begin{equation}
\mathcal{L}^{\mathrm{OPE}}(\tau)
=
\int_{t_q}^{\infty}
\!\!ds\,
e^{-s\tau} \,\rho^{\mathrm{OPE}}(s)
\label{eq:laplace_ope}
\end{equation}
On the hadronic side, the transformed correlator is
\begin{equation}
\mathcal{L}^{\mathrm{had}}(\tau)
=
\int_{t_{min}}^{\infty}
\!\!ds\,
e^{-s\tau} \,\rho^{\mathrm{had}}(s),
\label{eq:laplace_hadronic}
\end{equation}
where $t_{min}$ is the lowest physical threshold in the channel and
$\rho^{\mathrm{had}}(s)$ contains the contributions of all physical
states coupled to the tensor current. The exponential kernel
$e^{-s\tau}$ enhances the relative importance of the low-energy spectral
region while exponentially suppressing contributions from higher-energy
states and the continuum.

Quark--hadron duality states that the QCD and hadronic representations
describe the same correlation function after a suitable spectral smearing.
In the Laplace formulation, this matching is expressed as
\begin{equation}
\mathcal{L}^{\mathrm{OPE}}(\tau)
\simeq
\mathcal{L}^{\mathrm{had}}(\tau).
\label{eq:laplace_duality}
\end{equation}
Adopting the usual pole \(\oplus\) continuum parametrization,
\begin{equation}
\rho^{\mathrm{had}}(s)
=
\lambda_0^2 \,\delta(s-m_0^2)
+
\theta(s-t_c)\,\rho^{\mathrm{OPE}}(s),
\label{eq:pole_continuum_ansatz}
\end{equation}
where $m_0$ is the mass of the lowest state, $\lambda_0$ denotes its
coupling to the interpolating current, and $t_c$ is the effective
continuum threshold, the continuum contribution can be transferred to
the QCD side. The resulting continuum-subtracted Laplace sum rule is
\begin{equation}
\lambda_0^2 \,e^{-m_0^2\tau}
=
\int_{t_q}^{t_c}
ds\,
e^{-s\tau}\rho^{\mathrm{OPE}}(s).
\label{eq:continuum_subtracted_laplace_sr}
\end{equation}
More generally, the Laplace moments are defined as
\begin{equation}
\mathcal L_n(\tau,t_c)
=
\int_{t_q}^{t_c}
ds\,s^n e^{-s\tau}
\rho^{\mathrm{OPE}}(s).
\end{equation}
Within the parametrization in Eq.(\ref{eq:pole_continuum_ansatz}), 
the ground-state mass is extracted from the ratio of consecutive moments,
\begin{equation}
m_0^2(\tau,t_c)
=
\frac{\mathcal L_1(\tau,t_c)}
{\mathcal L_0(\tau,t_c)}.    
\end{equation}
The quantities \(\tau\) and \(t_c\) are not physical observables. 
In an exact calculation, hadronic masses and couplings would be
independent of them. Their residual influence in a sum rule 
calculations arises from the truncation of the 
perturbative series and the OPE, uncertainties in the QCD inputs, 
and the approximate description of excited states and the hadronic 
continuum. Consequently, the identification of an interval in $\tau$ 
where the OPE remains under control, the contribution of the 
low-energy spectrum is sufficiently enhanced, and the extracted 
observable is stable against variations of $\tau$ and the other 
auxiliary parameters constitutes an essential part of the method
\cite{Narison2004}. 

\subsection{\label{subsec:A} Borel Windows}
The \(\tau\) variable has a particularly important dual role. 
As \(\tau\) increases, high-energy states are more strongly 
suppressed by the exponential kernel, improving the relative 
sensitivity to the lowest-energy part of the spectrum. At the 
same time, power corrections associated with vacuum condensates 
become increasingly important, and the truncated OPE may cease to 
provide a controlled approximation. Conversely, at small \(\tau\), the 
short-distance expansion is generally better behaved, but the 
exponential suppression of excited states and continuum 
contributions becomes weaker. A reliable analysis must therefore 
be performed in an intermediate interval,
\[
\tau_{\min}\leq \tau\leq\tau_{\max}.
\]
This interval is conventionally referred to in the literature as 
the \emph{Borel window} and will henceforth be called the 
\emph{sum rule window} in this work.

Conventionally, the two boundaries of this interval are obtained from 
different requirements. The large-\(\tau\) boundary is constrained by 
OPE convergence, often by requiring the highest-dimensional term 
retained in the calculation to remain smaller than a prescribed 
fraction of the total OPE contribution. The small-\(\tau\) boundary 
is commonly determined by requiring a minimum pole contribution, or, 
more generally, sufficient sensitivity to the low-energy part of the 
spectral function. Inside the resulting interval, the hadronic 
observable is expected to exhibit stability with respect to \(\tau\). 
The continuum threshold \(t_c\) is varied simultaneously until a 
region of acceptable \(\tau\)-stability is found.

Although physically well motivated, this standard construction 
contains unavoidable elements of prescription. Different analyses may
adopt different percentages for OPE convergence and pole dominance, 
and a plateau is frequently selected through visual inspection. 
Moreover, \(\tau\) and \(t_c\) are correlated: changing the continuum 
threshold modifies both the relative continuum contribution and the 
dependence on the Laplace parameter $\tau$. The resulting stability 
can therefore depend nontrivially on the simultaneous choice of 
auxiliary parameters.

The limitations of stability arguments have also been demonstrated 
through toy models in which an apparently convincing plateau may 
arise even in the absence of a physical resonance \cite{Steele1997}.
Conversely, Monte Carlo analyses of QCD inputs have shown that a
quantitative treatment of uncertainties and parameter correlations 
can considerably improve the assessment of the predictive content 
of a sum rule \cite{Leinweber:1995fn}. These results motivate the 
development of reproducible procedures that combine theoretical 
constraints, parameter correlations, and local stability information
instead of relying predominantly on visual criteria.

In this work, we formulate the determination of the sum rule window 
as a constrained optimization problem. The objective is not merely 
to locate a single extremum of the extracted mass, but to identify 
an extended and connected domain in the space of auxiliary 
parameters in which the QCD calculation satisfies the adopted 
reliability conditions and the predicted hadronic observable 
exhibits minimal local sensitivity.

\subsection{\label{sec:B}Continuum Threshold (\(\bm{t_c}\))}
The effective threshold deserves particular attention. It is not, 
in general, a directly measurable quantity and should not be 
automatically identified with the squared mass of the first 
radial excitation. It represents an effective separation scale 
introduced by the pole \(\oplus\) continuum parametrization in 
Eq.(\ref{eq:pole_continuum_ansatz}).
Studies based on exactly solvable potential models have shown that 
the extraction of ground-state parameters can be strongly 
influenced by the continuum model and that apparently stable 
\(\tau\)-behavior does not necessarily guarantee an accurate result 
\cite{Lucha2007, Lucha2009, Lucha2010Thresholds}. In particular, the 
exact effective continuum threshold may depend on the Laplace 
parameter and on the correlator under consideration. These 
observations do not invalidate the conventional constant threshold 
approximation, but they show that stability must be interpreted as 
a necessary consistency condition rather than as an independent 
proof of reliability.

\subsection{\label{subsec:pole}Pole Dominance}
The extraction of a ground-state parameter requires a region of 
the sum rule window in which the low-energy part of the spectrum 
is sufficiently enhanced relative to excited states and the 
hadronic continuum. This requirement is usually referred to as 
\emph{pole dominance}. Its implementation, however, should not be 
reduced to a single fixed percentage. A reliable low-energy
selection must combine at least three complementary properties: 
i) stability with respect to the Laplace parameter $\tau$, 
ii) limited sensitivity to the continuum threshold $t_c$, and 
iii) compatibility with a spectral distribution dominated by a
single narrow state.

The meaning of $\tau$-stability can be made transparent by 
introducing the normalized Laplace-weighted spectral distribution
\begin{equation}
\widehat{\rho}(s,\tau,t_c)
=
\frac{e^{-s\tau}\rho(s)}
{\mathcal{L}_0(\tau,t_c)},
\qquad
\int_{t_q}^{t_c}
ds\,\widehat{\rho}(s,\tau,t_c)
=
1,
\label{eq:normalized_density}
\end{equation}
provided that the spectral density is positive in the integration 
domain. The squared mass, \(m_0^2\), can then be 
interpreted as the first moment of this distribution. Indeed,
if we evaluate the expectation value of \(s\)-variable,
\begin{equation}
\langle s\rangle_{\tau}
=
\int_{t_q}^{t_c}
ds\,s\,\widehat{\rho}(\tau,s;t_c)
=
\frac{\mathcal{L}_1}{\mathcal{L}_0}
=
m_{0}^2.
\label{eq:mass_as_expectation_value}
\end{equation}
Using
\begin{equation}
\frac{\partial\mathcal{L}_n}{\partial\tau}
=
-\mathcal{L}_{n+1},
\label{eq:moment_derivative_relation}
\end{equation}
one obtains
\begin{align}
\frac{\partial m_{0}^2}{\partial\tau}
&~=~
\frac{\mathcal{L}_1^2-\mathcal{L}_0\mathcal{L}_2}
     {\mathcal{L}_0^2}
~=~
-\Bigg[
\frac{\mathcal{L}_2}{\mathcal{L}_0}
-
\left(\frac{\mathcal{L}_1}{\mathcal{L}_0}\right)^2
\Bigg]
\nonumber\\
&~=
-\bigg[
\langle s^2\rangle_{\tau}
-
\langle s\rangle_{\tau}^2
\bigg]
=
-\,\sigma_s^2(\tau,t_c).
\label{eq:mass_derivative_variance}
\end{align}
Thus, for a positive spectral measure, the $\tau$ derivative of 
the extracted squared mass is the negative of the spectral 
variance in \(s\)-variable, \(\sigma_s^2\). A nearly stationary 
mass therefore indicates that the Laplace-weighted spectral 
strength is concentrated within a relatively narrow interval 
in $s$. Equivalently, defining
\begin{equation}
m_1^2(\tau,t_c)
=
\frac{\mathcal{L}_2(\tau,t_c)}
     {\mathcal{L}_1(\tau,t_c)},
\label{eq:first_excited_mass_ratio}
\end{equation}
Eq.~\eqref{eq:mass_derivative_variance} may be written as
\begin{equation}
\frac{\partial m_{0}^2}{\partial\tau}
=
m_{0}^2
\left(
m_{0}^2-m_1^2
\right).
\label{eq:mass_derivative_ratio_form}
\end{equation}
Now, as an example, consider an isolated narrow pole in the 
spectrum, with no other resonances and/or continuum contributions. 
For an isolated zero-width pole,
\begin{equation*}
\rho^{\mathrm{pole}}(s)
=
\lambda_A^2 \,\delta(s-m_A^2).
\end{equation*}
The first two moments are
\begin{equation*}
\mathcal{L}_0^{\mathrm{pole}}
=
\lambda_A^2 \,e^{-m_A^2\tau},
\qquad
\mathcal{L}_1^{\mathrm{pole}}
=
\lambda_A^2 \,m_A^2 \,e^{-m_A^2\tau},
\end{equation*}
and consequently
\begin{equation*}
m_{0}^2
=
\frac{\mathcal{L}_1^{\mathrm{pole}}}
     {\mathcal{L}_0^{\mathrm{pole}}} \,,
\qquad
\frac{\partial m_{0}^2}{\partial\tau}=0.
\end{equation*}
One finds that $m_0^2 = m_A^2$ and $\sigma_s^2=0$, so the mass estimator is exactly 
independent of $\tau$. According to Eq.(\ref{eq:mass_derivative_variance}), 
for a more general spectral distribution, a nonzero variance produces a 
negative slope. 
Increasing $\tau$ exponentially suppresses the higher-energy part 
of the spectrum and generally drives the spectral centroid toward 
the lowest-energy region coupled to the current.
A small spectral variance is nevertheless not sufficient to establish 
the presence of a single physical pole. Several nearly degenerate 
states, a narrow threshold enhancement, or a sufficiently localized
part of a continuous distribution may also produce a weak $\tau$ 
dependence. The variance relation should therefore be interpreted 
as a measure of spectral concentration under the Laplace kernel, 
rather than as independent evidence for a unique resonance.

The interpretation becomes particularly clear when two narrow states
contribute:
\begin{equation*}
\rho(s)
=
\lambda_A^2 \,\delta(s-m_A^2)
+
\lambda_B^2 \,\delta(s-m_B^2),
\qquad m_A<m_B.
\end{equation*}
In this case, the squared mass is a weighted average,
\begin{equation*}
m_{0}^2(\tau)
=
w_A(\tau)m_A^2 \,+\, w_B(\tau)m_B^2,
\end{equation*}
where:
\begin{equation*}
w_i(\tau)
=
\frac{\lambda_i^2 \,e^{-m_i^2\tau}}
{\lambda_A^2\,e^{-m_A^2\tau}
+\lambda_B^2\,e^{-m_B^2\tau}},
\qquad
w_A+w_B=1 \,.
\end{equation*}
Its derivative is
\begin{equation*}
\frac{\partial m_{0}^2}{\partial\tau}
=
-w_A \,w_B\left(m_B^2-m_A^2\right)^2.
\end{equation*}
Notice that the heavier state is exponentially suppressed as 
$\tau$ increases, 
\begin{equation*}
\frac{w_B}{w_A}
=
\frac{\lambda_B^2}{\lambda_A^2}
\,e^{-(m_B^2-m_A^2)\tau},
\end{equation*}
the heavier state is progressively suppressed as $\tau$ increases
(since \( w_B \to 0 \)), and \( m_0^2 \) mass approaches the squared mass 
of the lowest state $m_A^2$ (\( w_A \to 1 \)). The same reasoning 
extends to a continuous distribution: increasing $\tau$ reduces 
the relative importance of the high-energy tail and narrows the 
Laplace-weighted spectral distribution.

Consequently, a broad weighted spectral distribution produces a stronger
variation of the mass estimator with $\tau$, whereas a sufficiently
concentrated distribution leads to an approximately constant
$m_0^2(\tau,t_c)$ over the relevant $\tau$ interval.
Therefore, exact stationarity at an isolated value of $\tau$ 
should not be required. The physically relevant condition is a 
sufficiently small mass variation over an extended and connected 
interval that also satisfies OPE convergence, adequate continuum 
suppression, positivity of the relevant moments, and limited 
sensitivity to the continuum threshold $t_c$. A nearly flat region 
indicates that the spectral distribution selected by the
Laplace transform is sufficiently concentrated. It is compatible 
with the dominance of a resonance, but does not, by itself, 
constitute proof of the existence of a single pole.

\subsection{\label{subsec:D} 
Explicit Dependence of Spectral Density on $\bm \tau$}
The identities in Eqs.~\eqref{eq:moment_derivative_relation} and
\eqref{eq:mass_derivative_variance} assume that $\rho(s)$ has no 
explicit dependence on $\tau$. If the truncated QCD representation 
contains an explicit $\tau$-dependence, the derivative must be 
evaluated consistently. In that case, one may define
\begin{equation}
\widetilde{\mathcal{L}}_1(\tau,t_c)
\equiv
-\frac{\partial\mathcal{L}_0(\tau,t_c)}{\partial\tau},
\label{eq:generalized_first_moment}
\end{equation}
so that
\begin{equation}
m_{0}^2(\tau,t_c)
=
\frac{
\widetilde{\mathcal{L}}_1(\tau,t_c)
}{
\mathcal{L}_0(\tau,t_c)
}.
\label{eq:generalized_mass_ratio}
\end{equation}
For
\begin{equation}
\mathcal{L}_0(\tau,t_c)
=
\int_{t_q}^{t_c}
ds\,e^{-s\tau}\rho(\tau,s),
\end{equation}
the generalized first moment becomes
\begin{equation}
\widetilde{\mathcal{L}}_1(\tau,t_c)
=
\int_{t_q}^{t_c}
ds\,e^{-s\tau}
\left[
s\,\rho(\tau,s)
-
\frac{\partial\rho(\tau,s)}{\partial\tau}
\right].
\label{eq:generalized_first_moment_integral}
\end{equation}
Therefore, the simple identification
$\widetilde{\mathcal{L}}_1=\mathcal{L}_1$ and the variance 
relation in Eq.~\eqref{eq:mass_derivative_variance} cannot be 
used without accounting for the additional derivative terms. 
The stationarity condition should then be evaluated directly 
from the consistently differentiated mass ratio.
\\ \\ 

The optimization procedure proposed in this work is intended to reduce, as far
as possible, the dependence of QCD sum rule predictions on auxiliary and
phenomenological parameters, as well as on analysis choices that may otherwise
introduce a degree of arbitrariness. It should not be interpreted as an attempt 
to establish that this prescription is intrinsically better or worse than the 
conventional approaches commonly adopted in QCD sum rule studies. Rather, our 
purpose is to formulate a systematic and reproducible framework in which the 
admissible parameter domain is constrained by explicit physical and numerical
conditions, and the remaining parameter dependence can be identified, quantified, 
and incorporated into the final uncertainty.

The remainder of this paper is organized as follows. 
In Sec.~\ref{sec:entropy}, we define the OPE entropy and demonstrate how 
it constrains an initial domain in the \(\tau\)-parameter space, hereafter 
referred to as the \emph{Working Region}, within which the sum rule window 
is subsequently determined.
In Sec.~\ref{sec:formalism}, we introduce the tensor interpolating currents and 
the QCD Laplace sum rule formalism, using a fully charmed $cc\bar{c}\bar{c}$ 
state as a specific application through which the optimization procedure is 
illustrated. This section also discusses mixed interpolating currents and 
introduces the mixing parameters, including the parameter $\kappa$. Section~\ref{sec:optimization} develops the 
remaining stages of the optimization procedure, including the analysis of 
the mixing angle, the mapping of the continuum threshold $t_c$, the 
determination of its physically admissible interval, and the identification 
of the final sum rule window. In Sec.~\ref{sec:numerical-results}, we present 
the numerical mass prediction and its uncertainty analysis, followed by a 
comparison with previous theoretical calculations and with the structures 
observed in the di-$J/\psi$ spectrum.
Finally, Sec.~\ref{sec:conclusions} summarizes our conclusions regarding the
optimization procedure and discusses its role in controlling the dependence of
QCD sum rule results on the auxiliary parameters.

\section{OPE Entropy}
\label{sec:entropy}
In this section, we introduce the OPE entropy and show how it can be used 
to delimit an initial domain in the \(\tau\)-parameter space. This preliminary 
interval, the Working Region, provides the domain within which the sum rule 
window will subsequently be determined through the stability and optimization 
criteria discussed in the next section.

The OPE representation of the QCD Laplace moment in 
Eq.(\ref{eq:laplace_ope}) can be decomposed into a sum of \(N\) individual 
contributions,
\[
\mathcal{L}^{\mathrm{OPE}}(\tau)
=
\sum_{d=1}^N
\mathcal{L}_{0}^{(d)}(\tau),
\]
where the index \(d\) labels the perturbative term and the successive 
nonperturbative contributions associated with vacuum condensates of 
increasing dimension in the OPE. Although the truncated OPE is expected 
to provide a reliable description only within a limited range of \(\tau\), 
the relative importance of its individual terms changes continuously 
across this domain.
To quantify this redistribution, we define normalized OPE weights as
\begin{equation}
p_i(\tau)
=
\frac{
\left|
\mathcal{L}_{0}^{(i)}(\tau)
\right|
}{
\displaystyle
\sum_{d=1}^N
\left|
\mathcal{L}_{0}^{(d)}(\tau)
\right|
},
\qquad
\sum_{i=1}^N \,p_i=1.
\end{equation}
The absolute values are introduced because the purpose of the weights 
is to measure the magnitude of each OPE contribution rather than the 
cancellations that may occur in their sum. 

Inspired by Shannon’s information measure \cite{Shannon1948}, the 
normalized OPE entropy is defined as 
\begin{equation}
S_{\mathrm{OPE}}(\tau)
=
-\frac{1}{\ln N}
\sum_{i=1}^{N}
\,p_i(\tau)
\,\ln p_i(\tau).
\end{equation}
By construction,
\[
0\leq S_{\mathrm{OPE}}\leq 1.
\]
The entropy approaches zero when the Laplace moment is almost entirely 
dominated by a single OPE contribution. Conversely, it approaches unity when 
all terms contribute with comparable magnitudes. Thus, \(S_{\mathrm{OPE}}\) 
does not directly measure the numerical convergence of the OPE. Rather, it 
characterizes how the total QCD information is distributed among the 
perturbative and nonperturbative sectors.

This distinction provides a useful physical interpretation of the 
\(\tau\) dependence. At small \(\tau\), corresponding to the short-distance 
or high-energy regime, the exponential kernel \(e^{-s\tau}\) only weakly 
suppresses the high-energy part of the spectral distribution. The perturbative 
contribution then tends to dominate the Laplace moment, causing the entropy 
to approach small values. Although the OPE is generally well ordered in 
this regime, the continuum is insufficiently suppressed and the moment 
becomes less sensitive to the lowest hadronic state.

At large \(\tau\), corresponding to an enhanced sensitivity to lower-energy 
and long-distance dynamics, contributions from higher-dimensional condensates 
become progressively more important. The OPE information is consequently 
distributed among a larger number of terms, and the entropy increases. 
An excessively large entropy therefore signals a region in which several 
nonperturbative contributions have become simultaneously relevant and the 
reliability of the truncated expansion may begin to deteriorate.

These considerations motivate the introduction of a preliminary interval in the 
Laplace parameter, defined so as to exclude both limiting regimes. We denote 
this interval as the Working Region:
\begin{equation}
\Delta\tau_{\mathrm w} \,=\,
\tau_{\mathrm{sup}}-\tau_{\mathrm{inf}}.
\end{equation}
Since no unique entropy thresholds are fixed by first principles, the 
boundaries of this interval must be specified through an operational 
prescription. Generally, we can apply the following criteria:
\[
S_{\mathrm{OPE}}(\tau_{\mathrm{inf}})=0.10,
\qquad
S_{\mathrm{OPE}}(\tau_{\mathrm{sup}})=0.60.
\]
These threshold values are not universal and may be adjusted according to 
the specific sum rule under analysis, particularly they constitute 
prescriptions chosen to provide a sufficiently broad domain while avoiding 
the two regimes in which the sum rule loses sensitivity to the lowest state 
or the OPE becomes increasingly unreliable.

\begin{figure}[b]
    \centering
    \includegraphics[width=0.9\linewidth]{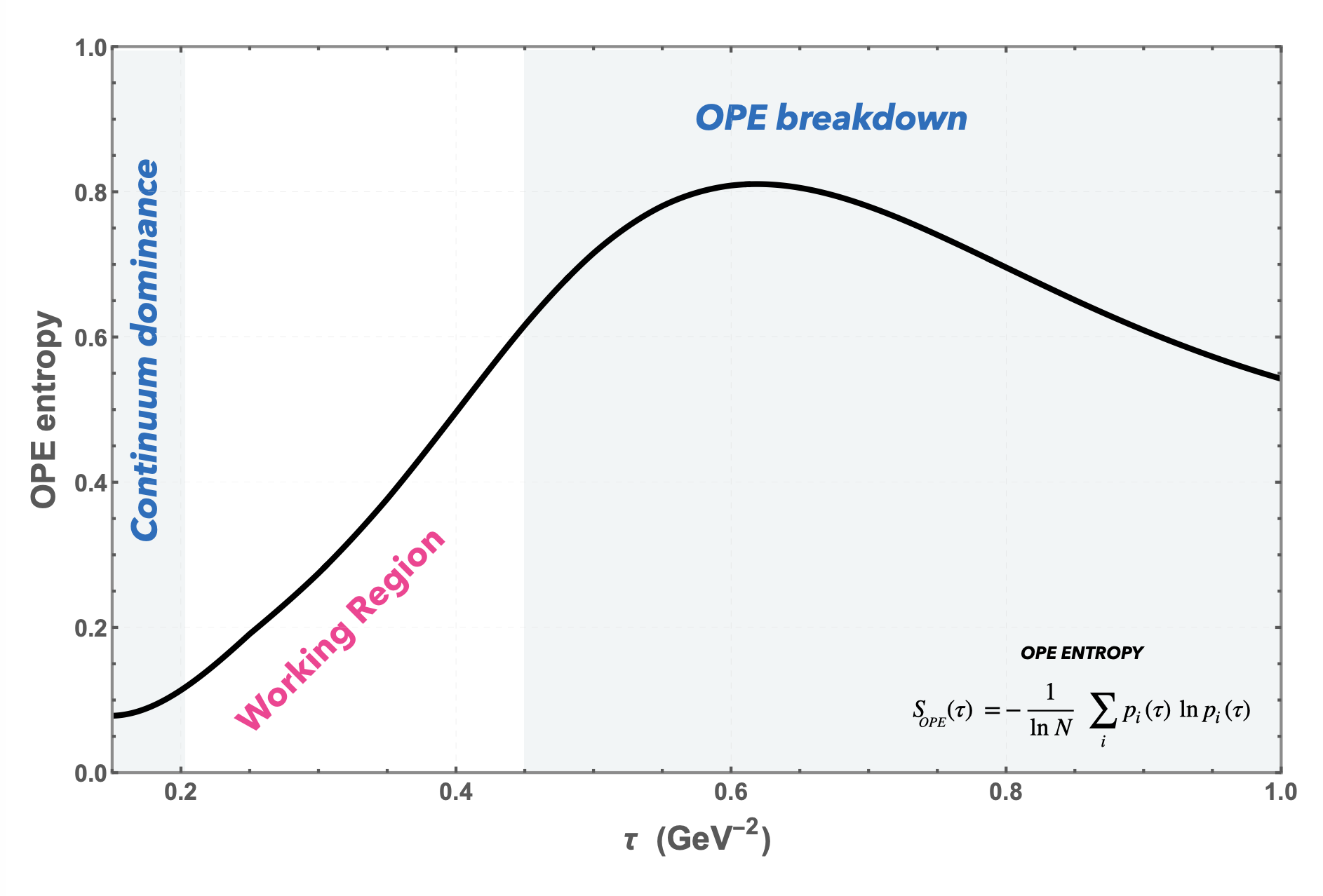}
    \caption{Schematic behavior of the normalized OPE entropy 
    \(S_{\mathrm{OPE}}(\tau)\) as a function of the Laplace parameter 
    \(\tau\).}
    \label{img:opeS}
\end{figure}

A schematic representation of the expected behavior of the OPE entropy 
is shown in Fig.\,\ref{img:opeS}, where the Working Region lies between 
the perturbative-dominated small-\(\tau\) regime and the large-\(\tau\) 
region associated with the deterioration of the truncated OPE.

This construction differs fundamentally from the Maximum Entropy Method 
previously applied to QCD sum rules \cite{Gubler2010}. In Bayesian 
maximum-entropy analyses, entropy is used as a regularizing functional for
reconstructing the hadronic spectral density from limited OPE information. By 
contrast, the OPE entropy introduced here acts entirely on the QCD side: 
it quantifies the relative distribution of the already calculated OPE 
contributions and is used as an auxiliary criterion in the determination of 
the fiducial \(\tau\)-domain. It neither reconstructs the spectral function 
nor replaces the pole \(\oplus\) continuum parametrization.

To the best of our knowledge, a normalized Shannon-type entropy constructed 
from the relative strengths of the OPE sectors has not previously been 
incorporated as a criterion for determining Borel windows in conventional 
QCD Laplace sum rules. The proposed framework combines this auxiliary function 
with mass-stability conditions and a correlated analysis of the continuum 
threshold and other free parameters. The purpose is to reduce the arbitrariness 
associated with fixed convergence percentages and visual plateau selection, 
while preserving the physical requirements underlying the traditional 
sum rule window.

\section{Tensor interpolating currents}
\label{sec:formalism}
In this section, we introduce the tensor interpolating currents and summarize 
the QCD Laplace sum rule formalism employed throughout the analysis. Although 
the optimization procedure developed in this work is general and can be applied 
to different hadronic channels, we use a fully charmed \(cc\bar c\bar c\) state 
with quantum numbers \(J^{PC}=2^{++}\) as a specific application. This system 
provides a particularly useful testing ground because different local operators 
can couple to the same physical state, allowing the role of interpolating current
mixing to be investigated explicitly.

For a fully charmed tensor tetraquark, one may construct currents corresponding 
to different internal color and Dirac configurations. In particular, compact 
diquark--antidiquark operators and meson--meson-like operators provide two 
complementary representations of the same \(J^{PC}=2^{++}\) channel.

A compact tensor current may be written in terms of an axial-vector diquark and 
an axial-vector antidiquark as
\begin{equation}
J_{\mu\nu}^{D}(x)
=
\epsilon_{abc}\epsilon_{dec}
\big[
c_a^{T}\!(x) C\gamma_\mu c_b (x)
\big]
\big[
\bar c_d(x) \gamma_\nu C \bar c_e^{T}\!(x)
\big]
\label{eq:compact_current}
\end{equation}
where \(a,b,c,d,e\) denote color indices, and \(C\) is the 
charge-conjugation matrix.

A meson--meson-like tensor current can be constructed from two 
color-singlet vector bilinears,
\begin{equation}
J_{\mu\nu}^{M}(x)
=
\big[
\bar c_a(x)\gamma_\mu c_a(x)
\big]
\big[
\bar c_b(x)\gamma_\nu c_b(x)
\big].
\label{eq:vector_vector_current}
\end{equation}
This operator has the quantum numbers of an \(S\)-wave \(J/\psi\,J/\psi\) 
configuration coupled to total spin \(J=2\). The labels \(D\) and \(M\) will 
be used below to distinguish the compact diquark--antidiquark and meson--meson 
structures, respectively.

It is important to emphasize that neither of the currents above couples 
exclusively to a single spin configuration. Being Lorentz tensors, they 
generally couple simultaneously to states with spin 0, 1, and 2. In particular, 
lower-spin components are not excluded at the operator level and are intrinsically 
present in the correlation function. For this reason, the isolation of the pure 
spin-2 contribution is performed only after constructing the two-point correlation 
function, where the appropriate Lorentz decomposition allows the separation of the 
different spin sectors.

The distinction between compact and meson--meson-like currents should not be 
interpreted as a strict separation between mutually exclusive physical 
configurations. Local operators with identical quantum numbers are generally 
related through Fierz rearrangements and may couple simultaneously to the same
hadronic states. Nevertheless, they can probe different components of the QCD 
correlation function with different strengths. Their relative contribution is 
therefore conveniently investigated by introducing a mixed interpolating current.

\subsection{Mixed currents and the parameter \(\kappa\)}
\label{subsec:mixed_currents}
We define the \(2^{++}\) tensor current as a linear combination of 
the two operators introduced above,
\begin{equation}
J_{\mu\nu}(x)
=
\cos\theta\,J_{\mu\nu}^{D}(x)
\,+\,
\kappa\sin\theta\,J_{\mu\nu}^{M}(x),
\label{eq:mixed_current}
\end{equation}
where \(\theta\) controls the relative orientation of the current in the space 
spanned by \(J_{\mu\nu}^{D}\) and \(J_{\mu\nu}^{M}\), while \(\kappa\) accounts 
for their relative normalization.

The parameter \(\theta\) should not, by itself, be interpreted as a directly 
observable hadronic mixing angle. Rather, it parametrizes a family of 
interpolating operators with the same quantum numbers. Different values of 
\(\theta\) correspond to different probes of the same spectral channel. The 
optimization procedure introduced below determines which combinations generate 
a QCD sum rule with improved OPE convergence, pole dominance, and mass stability.

Substituting Eq.~\eqref{eq:mixed_current}, the corresponding two-point 
correlation function is given by
\begin{align}
\!\!\Pi_{\mu\nu,\rho\sigma}(q)
={}&
\cos^{2}\theta\,
\Pi_{\mu\nu,\rho\sigma}^{DD}(q)
+
\kappa^{2}\sin^{2}\theta\,
\Pi_{\mu\nu,\rho\sigma}^{MM}(q)
\nonumber\\
& \!\!+
\kappa\sin\theta\cos\theta
\left[
\Pi_{\mu\nu,\rho\sigma}^{DM}(q)
+
\Pi_{\mu\nu,\rho\sigma}^{MD}(q)
\right],
\label{eq:mixed_correlator}
\end{align}
where
\begin{equation}
\Pi_{\mu\nu,\rho\sigma}^{ij}(q)
=
i\int d^{4}x\,e^{iq\cdot x}
\langle 0|
T\!\left[
J_{\mu\nu}^{i}(x)
J_{\rho\sigma}^{j\,\dagger}(0)
\right]
|0\rangle.
\label{eq:correlator_components}
\end{equation}
The first two contributions in Eq.~\eqref{eq:mixed_correlator} are the 
pure diagonal correlators, associated with the tensor diquark--antidiquark 
and meson--meson currents, respectively. The last two are the mixed, or 
interference, correlators and measure the overlap between the two operator 
structures at the level of the two-point function. Thus, \(\kappa\) rescales 
the pure \(MM\) contribution quadratically, while the interference contribution 
is rescaled linearly.

For \(\theta=0\), the correlator reduces to the pure compact 
diquark--antidiquark contribution, \(\Pi^{DD}_{\mu\nu,\rho\sigma}\). In the 
opposite limit, \(\theta=\pi/2\), it reduces to the pure meson--meson 
contribution, \(\Pi^{MM}_{\mu\nu,\rho\sigma}\). Intermediate values of \(\theta\)
generate a superposition of the two diagonal contributions together with
their interference terms.

The parameter \(\kappa\) plays a distinct role. The two basis currents may differ 
considerably in their intrinsic normalization, color factors, perturbative spectral 
strength, and/or coupling to the physical state. Consequently, equal numerical 
coefficients in Eq.~\eqref{eq:mixed_current} do not necessarily correspond to 
equal physical contributions to the correlation function. The parameter \(\kappa\) 
compensates for this relative normalization and allows the meson--meson 
component to be enhanced or suppressed independently of the angular parametrization.

A natural measure of the overall strength of an interpolating current is
provided by its zeroth Laplace moment,
\begin{eqnarray}
\mathcal{L}_{0}(\tau,t_c) \,\bigg|_{\theta \,=\, 0} &~\equiv~& 
\mathcal{L}_{0}^{DD}(\tau,t_c) ~\\  && \nonumber \\
\mathcal{L}_{0}(\tau,t_c) 
\,\bigg|_{\theta \,=\, \pi/2} &\equiv& 
\kappa^2 \,\mathcal{L}_{0}^{MM}(\tau,t_c)
\end{eqnarray}
On the phenomenological side, this quantity is schematically related to
the coupling of the current to the hadronic spectral function through
\begin{equation}
\mathcal{L}_{0}^i
\sim
\lambda_i^{2} \,e^{-m_i^2\tau}
+
\text{continuum contributions},
\end{equation}
where \(i = DD\) or \(MM\) and \(\lambda_i\) is the respective coupling constant.
Its magnitude therefore provides a convenient measure of the correlation
strength generated by each current. We consequently define the relative 
normalization by imposing
\begin{equation}
\mathcal{L}_{0}^{DD}(\tau,t_c)
=
\kappa^2(\tau,t_c)\,
\mathcal{L}_{0}^{MM}(\tau,t_c),
\label{eq:kappa_condition}
\end{equation}
which gives
\begin{equation}
\kappa(\tau,t_c)
=
\left[
\frac{
\mathcal{L}_{0}^{DD}(\tau,t_c)
}{
\mathcal{L}_{0}^{MM}(\tau,t_c)
}
\right]^{1/2}.
\label{eq:kappa_definition}
\end{equation}
This prescription places the two unmixed currents on a common correlation
scale before the mixing-angle optimization is performed. Then, for each 
fixed value of \(t_c\), the normalization factor \(\kappa(\tau,t_c)\) 
is evaluated throughout the corresponding working region and averaged over 
its \(\tau\) points. Repeating this procedure over the continuum-threshold 
scan produces the function \(\overline{\kappa}(t_c)\). Its graphical 
representation is then used to verify whether the relative normalization 
remains stable under changes of \(t_c\). Once a sufficiently flat domain 
is identified, the value adopted in the subsequent analysis is obtained by 
averaging \(\overline{\kappa}(t_c)\) over that stable interval, resulting in 
the constant value \(\bar\kappa_0\) which is used in the mixing-angle optimization.

Both \(\theta\) and \(\kappa\) will be treated as external parameters whose 
admissible values are constrained by the numerical and optimization criteria 
developed in the following sections.

\section{The Sum Rule Window}
\label{sec:optimization}
The Working Region obtained from the OPE entropy analysis provides an initial 
domain for the Laplace parameter \(\tau\), but it does not yet guarantee the 
stability of the extracted hadronic observable with respect to the mixing angle 
\(\theta\) and the continuum threshold \(t_c\). These auxiliary parameters must 
therefore be determined simultaneously, since an apparently stable behavior in 
one variable may be lost when the remaining parameters are varied. The purpose 
of the optimization procedure is thus to identify a common domain in 
\((\tau,t_c,\theta)\) for which the sum rule prediction is locally stationary 
and only weakly sensitive to the auxiliary parameters.

We use the zeroth-moment estimator
\begin{equation}
 m_0^2(\tau,t_c,\theta)
\end{equation}
as the quantity to be optimized. Since the mixing angle becomes an active
parameter of the optimization procedure, its dependence, previously implicit
through the mixed interpolating current and the corresponding Laplace moments,
will henceforth be displayed explicitly. The central stationarity condition is
\begin{equation}
 \frac{\partial m_0^2}{\partial\tau}\simeq 0.
 \label{eq:tau_stationarity}
\end{equation}
This key condition identifies regions where the predicted mass scale is minimally 
affected by variations of \(\tau\). Requiring it over an admissible range of 
\(t_c\) also provides a practical test of \(t_c\)-stability.

Once the initial Working Region,
\begin{equation}
    \tau_{\mathrm{inf}} \,\leq\, \tau \,\leq\, \tau_{\mathrm{sup}}
\end{equation}
has been determined, the mixing angle is scanned over the complete interval
\begin{equation}
 -\pi/2 ~\leq~ \theta ~\leq~ \pi/2.
\end{equation}

The initial \(t_c\) domain is selected from physical threshold considerations. 
In the present case, the meson--meson component of the interpolating current 
naturally introduces a two-hadron threshold determined by the sum of the 
constituent-meson masses. For a \(J/\psi J/\psi\)-like configuration, this 
threshold lies near \(2M_{J/\psi}\simeq 6.2~\mathrm{GeV}\). The initial scan may 
therefore be performed, for example, over
\begin{equation}
 6.5~\mathrm{GeV}
 ~\lesssim~ \sqrt{t_c}
 ~\lesssim~ 7.5~\mathrm{GeV},
\end{equation}
which is sufficiently broad to include the expected resonance region while 
remaining physically connected to the lowest relevant hadronic threshold.

Within the resulting three-dimensional parameter domain, we search for the 
sets \((\tau,t_c,\theta)\) satisfying Eq.~\eqref{eq:tau_stationarity}. This 
preliminary scan determines which mixing angles allow a stationary point to 
occur inside the Working Region and which continuum thresholds support such a 
solution. It therefore acts as a first consistency condition relating the 
\(\tau\)- and \(t_c\)-stability requirements.

A preliminary interval for the mixing angle is obtained by examining the two 
boundaries of the Working Region. First, at \(\tau=\tau_{\mathrm{sup}}\), the 
angle for which \(\partial_\tau m_0^2=0\) is determined using the largest 
value of \(t_c\) in the initial scan. The same condition is then imposed at
\(\tau=\tau_{\mathrm{inf}}\), using the smallest value of \(t_c\). 
The two resulting angles delimit an initial admissible interval for \(\theta\).
At this stage, the original Working Region may be slightly reduced whenever 
its boundary points fail to satisfy the combined stationarity requirements. 
Such a contraction is desirable because it removes marginal regions where the 
sum rule becomes especially sensitive to the auxiliary parameters.

To quantify the residual variation of \(m_0^2\) over the retained 
\(\tau\)-interval, we introduce the normalized residual
\begin{equation}
 \mathcal{R}(\theta,t_c)
 =
 \frac{
 \sqrt{
 \langle m_0^4 \rangle_{\tau} - \langle m_0^2 \rangle_\tau^2}}
 {
 \left|
 \langle m_0^2 \rangle_\tau
 \right|
 },
 \label{eq:normalized_residual}
\end{equation}
where the averages are taken over the retained Working Region.
This dimensionless quantity measures the relative dispersion of the mass 
estimator along the \(\tau\) variable. A small value of $\mathcal{R}$ 
therefore indicates that the mass estimator $m_0^2(\tau,t_c,\theta)$ varies 
only weakly over the retained $\tau$ interval, rather than being stationary 
only at an isolated value of $\tau$.

The optimization does not consist simply of selecting the absolute minimum 
of a single curve. Instead, we search for the lowest physically admissible 
value of \(t_c\) for which an angle inside the preliminary interval yields a 
sufficiently small residual. This prescription avoids unnecessarily large 
continuum thresholds, which could artificially enlarge the pole \(\oplus\) 
continuum domain while reducing the physical discrimination of the sum rule.

The scan is subsequently repeated with a finer resolution around the target
region selected in the first stage. Two complementary features are used in 
this refinement. First, the minima of the residual curves determine the 
continuum thresholds for which each angle produces the smallest variation of
\(m_0^2\) along the \(\tau\) direction. Second, the convergence and crossing 
of neighboring curves reveal a domain in which the residual becomes weakly
sensitive to variations of the mixing angle. The optimized \((t_c, \theta)\) 
domain is selected from the overlap of these two behaviors, rather than from 
the minimum of a single residual curve.

The refined analysis determines the admissible intervals
\begin{equation}
 \theta_{\mathrm{min}}
 ~\leq~ \theta ~\leq~
 \theta_{\mathrm{max}},
 \qquad
 t_c^{\mathrm{min}}
 ~\leq~ t_c ~\leq~
 t_c^{\mathrm{max}}.
 \label{eq:optimized_theta_tc}
\end{equation}
These intervals should not be interpreted as independent uncertainties around
two isolated optimal values. They define a correlated parameter domain in
which the residual remains small and the \(m_0^2(\tau, t_c, \theta)\) function 
exhibits weak sensitivity to both the current mixing and the continuum model.

The final sum rule window is determined by fixing the mixing angle at a
representative value (\(\theta^\ast\)) within its optimized interval, which may 
conveniently be chosen as its midpoint.
The derivative
\begin{equation*}
 \frac{\partial
 m_0^2(\tau,t_c, \theta^\ast)}
 {\partial\tau}
\end{equation*}
is then evaluated at the lower and upper boundaries of the optimized
\(t_c\) interval. The zero of each limiting curve determines the position of 
the stationary point associated with one boundary of the admissible \(t_c\) 
domain. The interval enclosed by these zero crossings defines the region where 
the stationary points remain inside the accepted \(\tau\)-domain as \(t_c\) 
varies continuously between its optimized limits.

Each stage of the optimization imposes a distinct physical requirement: the
OPE entropy controls the distribution of the QCD contributions, the residual
quantifies the variation of the extracted mass scale, the angular scan tests
its dependence on the interpolating-current composition, and the threshold
scan assesses the separation between the lowest state and the higher-energy
spectrum.
The procedure therefore does not merely search for a numerical minimum.
Rather, it identifies a finite parameter domain in which the hadronic
prediction is simultaneously compatible with OPE reliability, spectral
separation, and local stationarity. In this way, the initial Working Region is
converted into the final sum rule window:
\begin{equation}
 \tau_{\mathrm{min}}
 ~\leq~ \tau ~\leq~
 \tau_{\mathrm{max}}.
 \label{eq:final_sr_window}
\end{equation}

\section{Numerical Results and Applications}
\label{sec:numerical-results}
We now apply the optimization procedure developed in the previous sections to
the fully charmed \(cc\bar c\bar c\) system with quantum numbers
\(J^{PC}=2^{++}\). The state is interpolated by the mixed tensor current
introduced in Sec.~\ref{sec:formalism}, which combines compact
diquark--antidiquark and color-singlet meson--meson operators. The numerical
analysis is performed inside the optimized sum rule window selected by the
combined OPE, pole-dominance, and mass-stationarity constraints.

\subsection{QCD Input Parameters}
\label{subsec:qcd-inputs}
Since the interpolating currents contain only charm-quark fields, the
nonperturbative sector of the OPE involves exclusively gluonic vacuum
condensates. We adopt the QCD input set employed in the NLO Laplace sum-rule
analysis of fully heavy molecules and tetraquarks presented in
Ref.~\cite{Albuquerque2020FullyHeavy}. The numerical values are
\begin{align*}
 \alpha_s(M_Z)
   &=0.1181\pm0.0016\pm0.0003,
 \\
 \overline m_c\!\left(\overline m_c\right)
   &=(1.286\pm0.016)\ {\rm GeV},
 \\
 \left\langle\alpha_sG^2\right\rangle
   &=(6.35\pm0.35)\times10^{-2}\ {\rm GeV}^{4},
 \\
 \left\langle g_s^3 G^3 \right\rangle
   &= (8.2\pm2.0)~{\rm GeV}^{2}
      \left\langle\alpha_s G^2\right\rangle .
\end{align*}
The strong coupling is obtained from heavy-quarkonium mass-splitting
in Laplace sum rules \cite{Narison2018QCDParameters}, while the running
charm-quark mass follows from sum rule analyses of the \(J/\psi\)
and \(B_c\) systems \cite{Narison2018HeavyQuarkMasses,
Narison2020HeavyQuarkMasses}. The dimension-four gluon condensate is taken
from the correlated QCD sum-rule analysis of Ref.~\cite{Narison2018QCDParameters}. 
For the dimension-six condensate, we use the value from charmonium sum rule 
analyses \cite{Narison2010GluonCondensates,Narison2011LightQuarks,
Narison2012GluonCondensates}. These inputs and their implementation in
fully heavy four-quark systems follow Ref.~\cite{Albuquerque2020FullyHeavy}.

The coupling \(\alpha_s(\mu)\) and the running mass
\(\overline m_c(\mu)\) are evolved consistently to the subtraction scale
using the renormalization-group equations for \(n_f=4\). For our
analysis, we adopt
\begin{equation}
 \mu=(4.5\pm0.2)~{\rm GeV}.
 \label{eq:mu-stability}
\end{equation}
This choice follows the \(\mu\)-stability analysis of fully charmed
four-quark systems in Ref.~\cite{Albuquerque2020FullyHeavy}, where the
corresponding observables exhibit minimum sensitivity around this value.
Compatible optimal scales were also obtained in independent heavy-quark
sum-rule analyses
\cite{Narison2015HigherOrders,Narison2020HeavyQuarkMasses,
Narison2020BcLike}. We use \(\mu=4.5~{\rm GeV}\) as the central value and
vary it within the interval in Eq.~\eqref{eq:mu-stability} to estimate the
residual renormalization-scale uncertainty.

The complete leading-order spectral densities of the tensor correlator are
used, including both factorizable and nonfactorizable contractions. The
next-to-leading-order perturbative correction is incorporated through the
factorization approximation of
Refs.~\cite{Albuquerque2018HigherOrders,
Albuquerque2020FullyHeavy}, in which the four-quark spectral density is
expressed as a convolution of heavy-quark bilinear spectral functions
\cite{Pich1985Convolution,Narison1994Convolution}. 

The OPE is truncated at dimension six and contains the perturbative,
\(\langle\alpha_sG^2\rangle\), and
\(\langle g_s^3G^3\rangle\) contributions. The uncertainties associated
with uncalculated higher perturbative orders and higher-dimensional gluonic
operators are included in the systematic error.

\subsection{Mass Prediction}
\label{subsec:optimization_results}

\begin{figure}[b]
    \centering
    \includegraphics[width=0.8\linewidth]{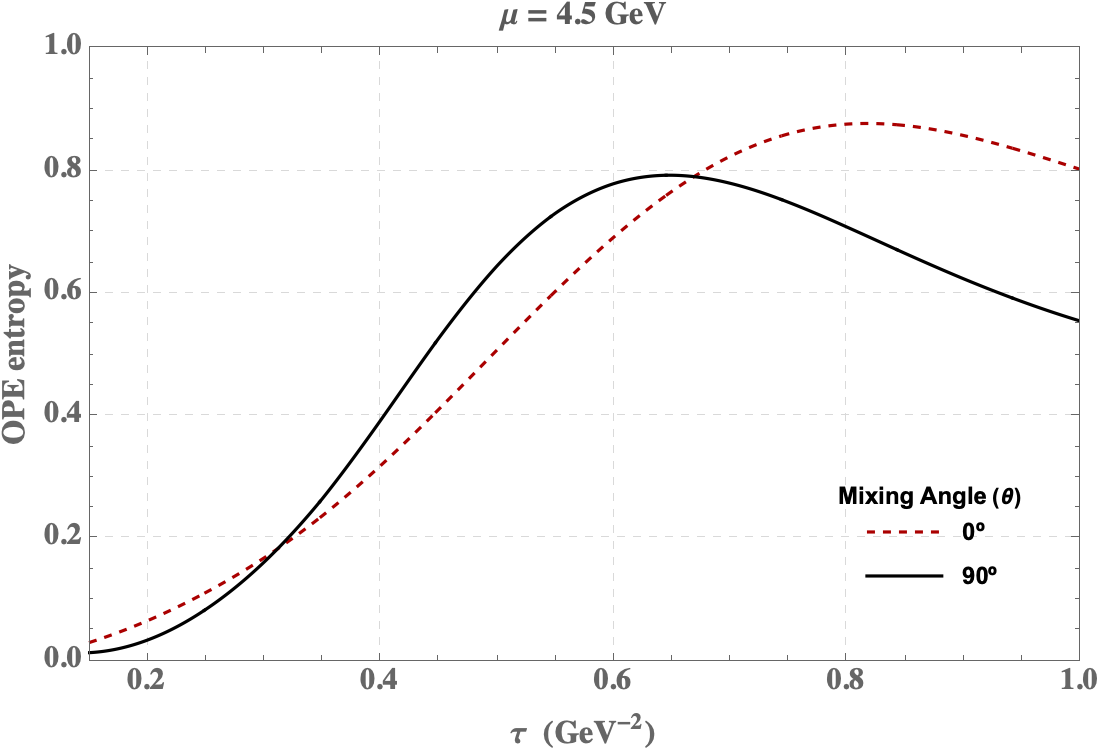}
    \caption{Normalized OPE entropy used to determine the initial 
    Working Region, at \(\mu=4.5~\mathrm{GeV}\). 
    The red dashed and black solid curves correspond, respectively, to the 
    pure diquark--antidiquark \((\theta=0^\circ)\) and pure meson--meson 
    \((\theta=90^\circ)\) configurations in the current.}
    \label{fig:entropy_results}
\end{figure}
For a mixed interpolating current, the initial Working Region is determined
from the OPE entropy behavior of the two pure-current limits. These limits
provide complementary information on the convergence of the corresponding
OPE. Since the physical current contains both components, neither 
pure-current entropy curve is used independently to define
the final interval. Instead, an intermediate criterion is adopted to construct
a representative Working Region that remains compatible with the convergence
properties of both configurations.

Figure~\ref{fig:entropy_results} shows the normalized OPE entropy as a function
of \(\tau\)-parameter for the pure components of the mixed current. The
selected Working Region lies within the common domain in which the OPE remains
sufficiently controlled for both configurations. This interval provides the
initial \(\tau\) domain on which the subsequent stability analysis is
performed: \(0.25 \,\leq\, \tau \,\leq\, 0.51 ~\mathrm{GeV^{-2}}\).

\begin{figure}[t]
    \centering
    \includegraphics[width=0.8\linewidth]{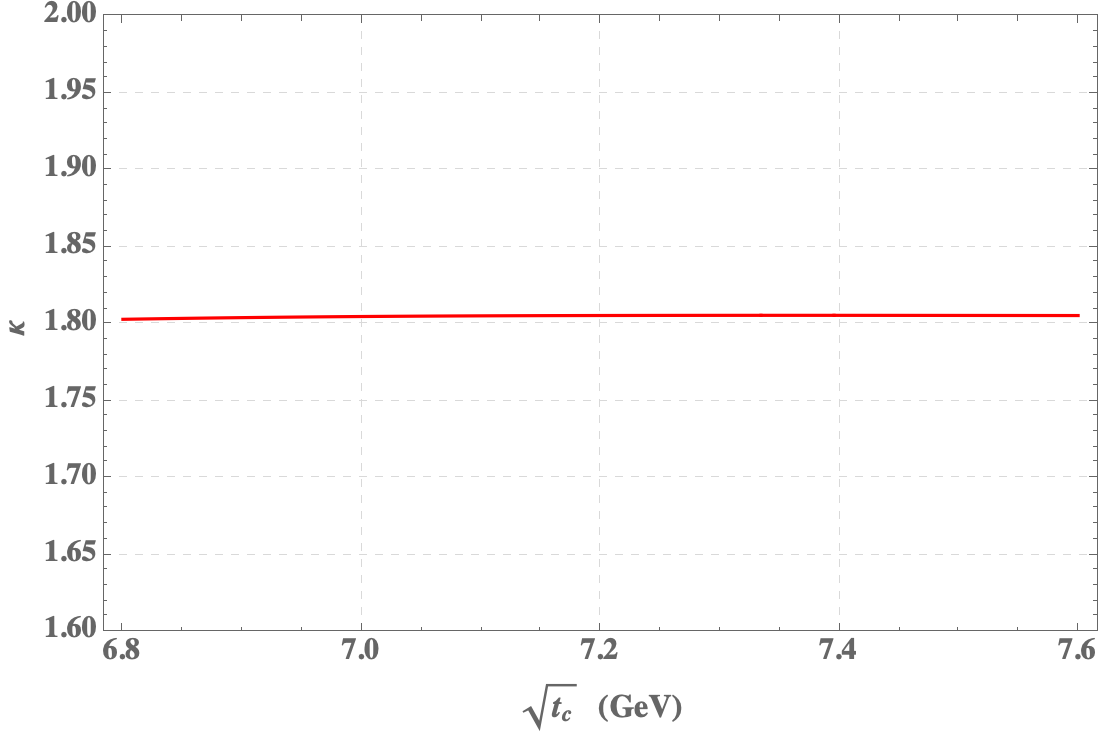}
    \caption{Dependence of the current normalization parameter \(\kappa\) 
    on the continuum threshold within the working region.}
    \label{fig:kappa_tc}
\end{figure}
Once the Working Region has been established, the relative normalization of
the two current components is fixed through the parameter \(\kappa\).
Figure~\ref{fig:kappa_tc} displays the resulting dependence of \(\kappa\) on
the continuum threshold. Its smooth variation over the admissible
threshold domain indicates that the normalization prescription remains stable
throughout the region relevant to the sum rule,
\[
\kappa \,=\, 1.81 \pm 0.13
\]

\begin{figure}
    \centering
    \includegraphics[width=0.9\linewidth]{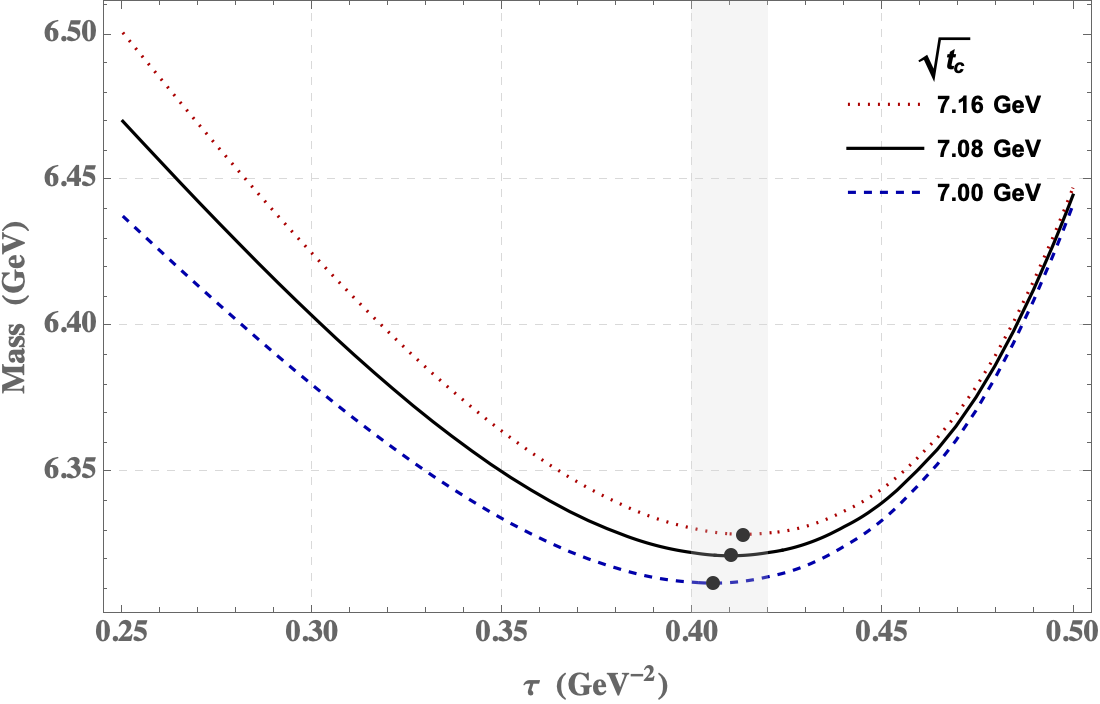}
    \caption{The mass \(M^{2^{++}}_{cc\bar{c}\bar{c}}\) at NLO as function of 
    \(\tau\), for diﬀerent values of \(t_c\), for \(\mu = 4.5\) GeV. The
    shadowed region delimits the sum rule window.}
    \label{fig:mass-window}
\end{figure}

The simultaneous optimization in \(\theta\), \(t_c\), and \(\tau\) is carried
out according to the stationarity and residual criteria introduced in
Sec.~\ref{sec:optimization}. In the following, the intermediate scans in
\(\theta\) and \(t_c\) are not shown explicitly. Their role is to retain the
correlated parameter domain for which the stationary behavior of
\(m_0^2(\tau,t_c,\theta)\) is preserved under variations of both the \(t_c\)-
and \(\tau\)- parameters. For the mixed tensor current, the optimal mixing 
angle is found to be
\begin{equation}
 \theta = 17.2^\circ \pm 0.4^\circ ,
\end{equation}
while the continuum threshold and the \(\tau\)-parameter are restricted to 
\begin{eqnarray}
 \sqrt{t_c} &=& 7.08 \pm 0.08 ~{\rm GeV},
 \label{eq:tc-optimal-range}
 \\
 \tau &=& 0.410 \pm 0.010 ~{\rm GeV}^{-2}.
 \label{eq:tau-optimal-range}
\end{eqnarray}
Within this domain, the mass value remains stationary over a
finite interval, while the OPE and pole-dominance constraints continue to
be simultaneously satisfied. The Fig.~\ref{fig:mass-window} displays the
resulting mass inside the optimized sum rule window.

The uncertainty is obtained by varying the QCD input parameters and the
external sum rule variables within their allowed intervals. 
The perturbative and OPE truncation uncertainties are also taken
into account. The individual contributions are combined in quadrature,
leading to
\begin{equation}
 M_{cc\bar c\bar c}^{\,2^{++}}
 =
 \left(6.32 \pm 0.11\right)\ {\rm GeV}.
 \label{eq:final-mass}
\end{equation}
This result corresponds to the lowest fully charmed tensor state coupled to
the optimized mixed interpolating current. Since the optimal value
\(\theta=17.2^\circ\) differs from both limiting cases, the extracted state
contains contributions from the compact diquark--antidiquark and
meson--meson operator structures. Nevertheless, the mass prediction alone
does not provide an unambiguous determination of the internal composition
of the state.

The present calculation is intended as a concrete application of the
optimization procedure developed in this work. A more detailed analysis of
the \(cc\bar c\bar c\), \(J^{PC}=2^{++}\) state, including its current
dependence, internal structure, spectral densities, 
coupling to the di-\(J/\psi\) channel, and possible decay mechanisms, 
will be presented in a dedicated forthcoming publication.

We now place the prediction in a broader phenomenological context by
comparing Eq.~\eqref{eq:final-mass} with previous theoretical calculations
and with the structures observed in the di-\(J/\psi\) invariant-mass
spectrum. This comparison will allow us to identify which experimental
enhancements are compatible with the optimized mass region, while avoiding
a definitive spectroscopic assignment based solely on the mass.

\subsection{Comparison with the di-\(J/\psi\) Spectrum}
\label{subsec:experimental-comparison}

The predicted mass in Eq.~\eqref{eq:final-mass} lies in the low-mass region
of the di-\(J/\psi\) invariant-mass spectrum, close to the kinematic
threshold. This region has received different phenomenological
interpretations as the experimental data and fitting strategies have
evolved.

\begin{figure}[b]
    \centering
    \includegraphics[width=1\linewidth]{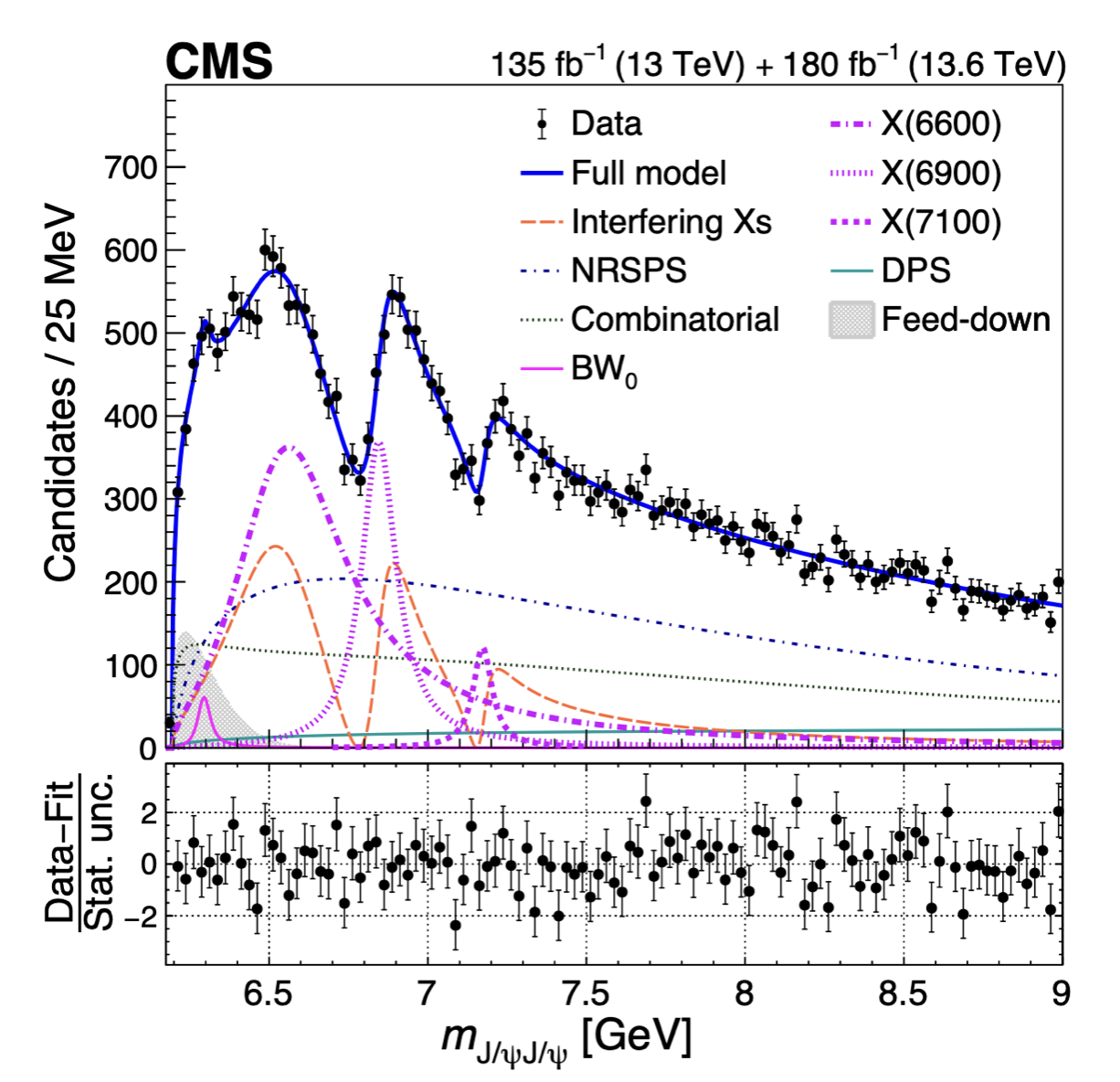}
    \caption{CMS fit to the di-\(J/\psi\) invariant-mass spectrum, 
    including the near-threshold \(\mathrm{BW}_0\) enhancement and 
    the \(X(6600)\), \(X(6900)\), and \(X(7100)\) structures. 
    Reproduced from Ref.\cite{CMS2026AllCharm}.}
    \label{fig:cms}
\end{figure}
The first detailed analysis by the LHCb Collaboration revealed a broad
enhancement immediately above the di-\(J/\psi\) threshold, extending
approximately from \(6.2\) to \(6.8~{\rm GeV}\), together with the narrower
structure around \(6.9~{\rm GeV}\)
\cite{LHCb2020DiJpsi}. The low-mass enhancement could be described either
by two Breit--Wigner components or by a resonance interfering with the
nonresonant single-parton-scattering contribution. Consequently, its
physical origin could not be uniquely established from the available data.

The subsequent ATLAS analysis also favored a resonant description of the
low-mass region. In the model containing three interfering resonances, the
lowest structure was found at
\begin{equation}
 M_0^{\rm ATLAS}
 =
 (6.41^{+0.08}_{-0.03})\ {\rm GeV},
 \label{eq:atlas-low-mass}
\end{equation}
followed by structures near \(6.63\) and \(6.86~{\rm GeV}\)
\cite{ATLAS2023DiCharmonium}. Although alternative contributions from
feed-down processes and threshold effects could not be excluded, this
analysis demonstrated that the low-mass enhancement can be consistently
modeled with a resonant component.

More recently, the CMS Collaboration separated the structures
\(X(6600)\), \(X(6900)\), and \(X(7100)\) from an additional near-threshold
component denoted by \(\mathrm{BW}_0\)
\cite{CMS2026AllCharm}. The three \(X\) structures were established with
significances above \(5\sigma\) and found evidence for mutual interference 
among them, with masses (see Fig. \ref{fig:cms})
\begin{align}
 M_{X(6600)}
   &=6593^{+15}_{-14}\pm25~{\rm MeV},
 \\
 M_{X(6900)}
   &=6847\pm10\pm15~{\rm MeV},
 \\
 M_{X(7100)}
   &=7173^{+9}_{-10}\pm13~{\rm MeV}.
\end{align}
In the nominal CMS fit, the near-threshold enhancement is represented by a
noninterfering Breit--Wigner function and included among the background
components. The CMS analysis also emphasizes that an unexplained excess 
of \(J/\psi\) pairs remains immediately above threshold, as previously 
observed by LHCb and ATLAS.

Our result, given by Eq.\,(\ref{eq:final-mass}), falls inside this broad 
near-threshold region and is consistent, within uncertainties, with the 
lowest resonant component obtained in the ATLAS fit.
It is, however, below the central mass of the established \(X(6600)\)
structure. The most direct experimental region associated with the present
prediction is therefore the enhancement currently parametrized by
\(\mathrm{BW}_0\) in the CMS analysis.

The mass compatibility indicates that a low-lying \(2^{++}\) contribution 
is not excluded in the near-threshold region. Nevertheless, the present 
sum rule prediction cannot distinguish a genuine resonance from threshold 
effects, feed-down contributions, nonresonant production, or overlapping 
amplitudes. A direct identification would require an experimental amplitude 
analysis capable of isolating a distinct near-threshold component and 
determining its quantum numbers.

Thus, the present result should be regarded as a theoretical mass prediction 
compatible with the unresolved near-threshold region of the di-\(J/\psi)\) 
spectrum, rather than as an identification of a specific experimental component. 
Whether this region contains an additional fully charmed tensor resonance 
remains an open experimental question. 
The detailed study of its decay into the di-\(J/\psi\)
channel and of observables capable of distinguishing a \(2^{++}\) resonance
from nonresonant mechanisms will be addressed in a dedicated forthcoming
work.

\section{Conclusions}
\label{sec:conclusions}
We have formulated the determination of the QCD Laplace sum rule window as a 
constrained optimization problem. The central purpose of the framework is 
to reduce the arbitrariness associated with fixed convergence percentages, 
isolated stability points, and the visual selection of mass plateaus. Rather 
than identifying a single optimal point, the procedure searches for an extended 
and connected parameter domain in which the OPE remains under control, the 
low-energy spectral region is sufficiently enhanced, and the extracted 
observable exhibits weak sensitivity to the auxiliary parameters.

As a concrete application, we considered the lowest fully charmed 
tetraquark state with $J^{PC}=2^{++}$, interpolated by a linear 
combination of compact diquark--antidiquark and color-singlet meson--meson 
currents. Within the sum rule window, the mass remains stationary while 
the adopted OPE and low-energy selection criteria are simultaneously satisfied. 
Including the uncertainties from the QCD input parameters, the renormalization 
scale, the external sum rule variables, and the perturbative and 
OPE truncations, we obtain
\begin{equation*}
M_{c\bar c c\bar c}^{2^{++}}
=
(6.32\pm0.11)~\mathrm{GeV}.
\end{equation*}

This prediction lies in the near-threshold region of the di-$J/\psi$ 
invariant-mass spectrum. It is compatible, within uncertainties, 
with the lowest resonant component obtained by ATLAS and points most 
directly to the enhancement parametrized as \(BW_0\) in the recent CMS analysis. 
The result therefore supports the possibility that the near-threshold 
region contains a low-lying fully charmed tensor contribution. Nevertheless, 
mass compatibility alone is insufficient for a definitive spectroscopic 
assignment. Threshold effects, feed-down mechanisms, nonresonant production, 
interference, and overlapping amplitudes may also contribute in this region. 
An experimental determination of the quantum numbers, together with a detailed 
analysis of the coupling and decay into the di-$J/\psi$ channel, 
remains essential.

The present application should primarily be regarded as an illustration 
of the optimization strategy. The OPE entropy thresholds and the 
detailed numerical implementation may require adaptation to the spectral 
properties and OPE structure of each channel. The general framework, however, 
is not restricted to fully heavy tetraquarks or to mixed currents. It can 
be applied to conventional hadrons, multiquark systems, hybrids, glueballs, 
and other channels in which several auxiliary parameters and competing 
interpolating operators are present.

In this sense, the proposed method complements rather than replaces the 
conventional QCD sum rule criteria. Its main advantage is to organize them 
within a reproducible procedure that explicitly exposes parameter correlations, 
identifies finite regions of minimal sensitivity, and incorporates the 
remaining auxiliary-parameter dependence into the quoted theoretical 
uncertainty.

\begin{acknowledgments}
The author would like to thank S. Narison, A. Rabemananjara, 
and D. Rabetiarivony for stimulating discussions and valuable 
suggestions that contributed to the development of this work. 
The main results were presented at the 29th High-Energy Physics 
International Conference in QCD (QCD26), held in Montpellier, France.
\end{acknowledgments}

\bibliography{apssamp}

\end{document}